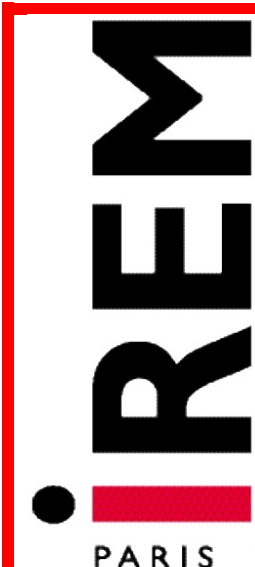
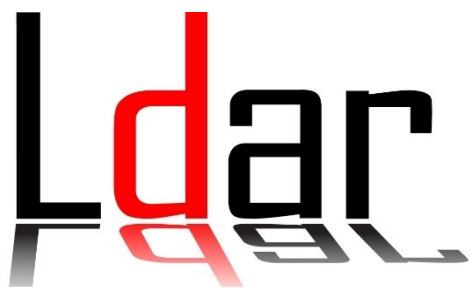

Laboratoire de didactique André Revuz
Mathématiques • Physique • Chimie

# Cahiers du laboratoire de didactique André Revuz n°18

Mai 2017

# Pour une théorie de l'activité en didactique des mathématiques

Un résumé des fondements partagés,
des développements récents et des perspectives

Par Maha Abboud-Blanchard, Aline Robert, Janine Rogalski et Fabrice Vandebrouck

**Coordonnées de l'IREM**

Pour venir à l'IREM (il est possible de consulter et d'acheter les publications sur place):
Université Paris-Diderot, Bâtiment Sophie-Germain,
8 place Aurélie Nemours (sur l'avenue de France), huitième étage,
 75013 Paris  13ème arrondissement
(métro 14 -Bibliothèque François Mitterrand ou tramway ligne T3a – Avenue de france )

**Nous Contacter**

Pour téléphoner: 01 57 27 91 93

Pour écrire à l'IREM concernant les publications:

*par voie postale:*
**Locufier Nadine**
**IREM de Paris – Case 7018**
**Université Paris Diderot**
**75205 Paris cedex 13**

*par voie électronique:*
**nlocufier@irem.univ-paris-diderot.fr**

La liste des publications de l'IREM est mise à jour sur notre site web :

**http://www.irem.univ-paris-diderot.fr/**   (en bas à gauche de la page d'accueil)

Pour rester informé:

 inscription à la liste de diffusion de l'IREM de Paris également sur le site de l'IREM

# Pour une théorie de l'activité en didactique des mathématiques

Un résumé des fondements partagés,

des développements récents et

des perspectives

M. Abboud

A. Robert

J. Rogalski

F. Vandebrouck

Ce texte ne vise pas une initiation à une théorie didactique de l'activité : c'est plutôt un récapitulatif condensé de ce qui est partagé par un certain nombre de chercheurs, et développé dans des directions variées.

Ces quelques pages présentent ainsi de manière succincte la manière dont le cadre de la Théorie de l'Activité, adopté depuis plusieurs années par des chercheurs en didactique des mathématiques, a été adapté pour étudier les apprentissages scolaires des mathématiques en relation avec l'enseignement que les élèves reçoivent, ainsi que les pratiques de leurs enseignants.

Des traits généraux communs de la méthodologie, qu'ensuite chacun a travaillée et continue de développer selon ses propres objets de recherche, permettent de préciser différents emprunts aux théories de l'apprentissage.

Quelques résultats sont esquissés, ainsi que des difficultés, des perspectives et des questionnements.

Mots clefs : théorie de l'activité, didactique des mathématiques

Nos recherches sont pilotées par la compréhension de l'apprentissage mathématique des élèves dans le cadre de l'enseignement qu'ils reçoivent à l'école (primaire, secondaire, début d'université).

Apprentissage des élèves

Pour attaquer le problème, nous avons choisi d'étudier les activités (mathématiques) des élèves en classe – ce qu'ils font (ou non), disent (ou non), écrivent (ou non), même si nous ne pouvons en recueillir que des traces, ce qu'ils pensent restant inobservable. **Cela traduit notre inscription dans la Théorie de l'Activité, adaptée à étudier des sujets en situation, en distinguant tâches et activités, en nous centrant sur les apprentissages. Cela inclut, plus précisément, le schéma de double régulation de Leplat (1997). Il y a là un emprunt « global » orientant et pilotant nos recherches.**

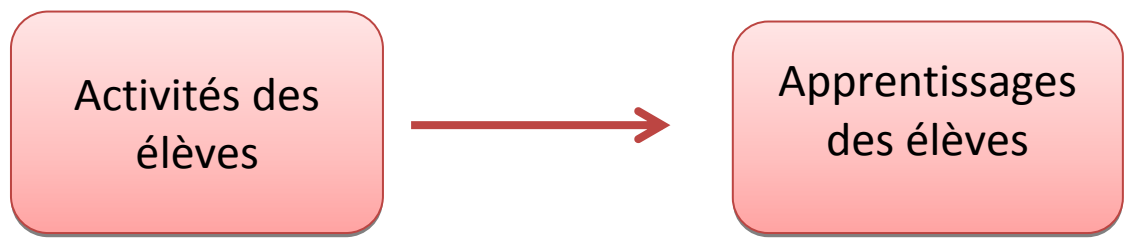

Ces activités des élèves sont provoquées pour une large part par les activités de l'enseignant (en classe).

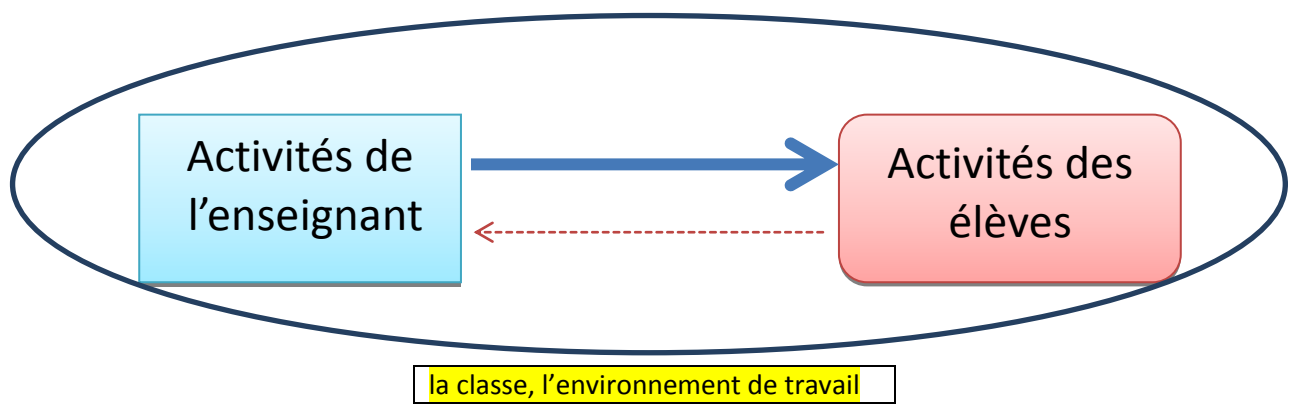

la classe, l'environnement de travail

Nos objets de recherche sont ainsi d'une part « la flèche » qui va des activités aux apprentissages des élèves, même si c'est lié à des hypothèses globales plus qu'à des résultats précis, et d'autre part, en étroite relation, les deux flèches (inégales) entre activités des enseignants et activités des élèves en classe. **Dans un but de diagnostic ou d'élaboration de ce qui se passe (peut se passer) en classe.**

Notre démarche est donc à la fois **expérimentale** (observations, recueil de données) **et théorique** (analyses de données, voire mise en place de ressources, élaboration de méthodologies adaptées), avec éventuellement des dialectiques entre phases.

Mais comment accéder aux activités des élèves et des enseignants ?

L'idée générale, pour les élèves, est d'analyser les tâches proposées en classe (localement et dans leur ensemble), et de comparer les activités des élèves attendues, en termes de mises en fonctionnement des connaissances, et leurs activités possibles, déduites des déroulements effectifs analysés. Par exemple dans certains cas l'enseignant aide les élèves, en réduisant la tâche et donc l'activité, ou alors il développe un discours qui permet une certaine généralisation de ce qu'ont fait les élèves, s'appuyant sur leur ZPD. Dans les phases d'exposition des connaissances, où les activités des élèves sont particulièrement inaccessibles, on les déduit surtout de l'étude fine du discours de l'enseignant. Par connaissances précisons que nous entendons des connaissances opérationnelles, c'est-à-dire non pas seulement le texte du savoir mais aussi sa mise en œuvre.

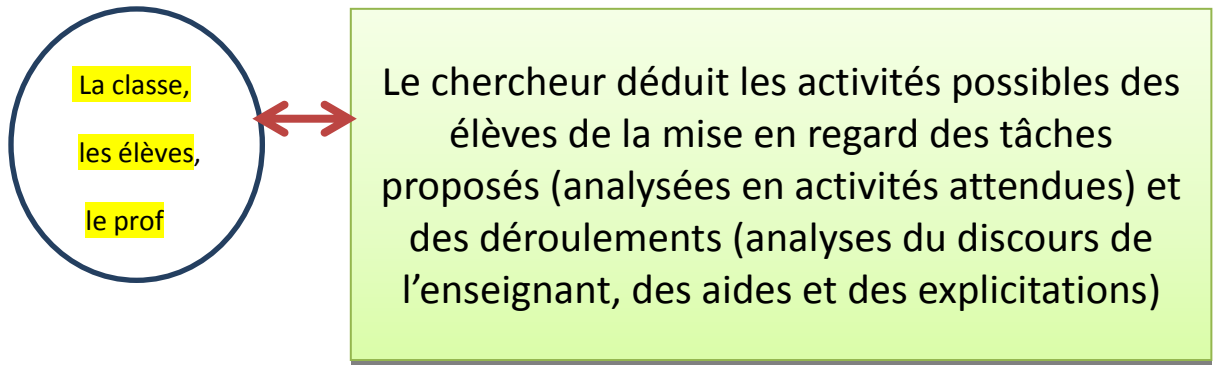

(*On l'aura compris, dans les bulles le fond bleu est associé aux élèves, le rouge aux professeurs, le vert (et le violet) aux chercheurs et le jaune au reste*.)

Toutes nos analyses mettent en jeu, à titre de référence mathématique, à établir très tôt dans nos recherches, ce que nous appelons le relief sur les notions à enseigner : il s'agit d'une étude croisant les aspects épistémo-mathématiques, curriculaires et cognitifs (difficultés déjà répertoriées par exemple) des notions. C'est ce croisement qui amène à distinguer des types de notions, par exemple les notions associées à un formalisme nouveau, généralisateur et unificateur, loin de ce que les élèves peuvent déjà connaître, étant supposées particulièrement difficiles à introduire.

Soulignons ici que la TSD et, à sa façon la TAD (voire la TACD) disposent d'études du même type (cf. notre entrée commune par les contenus à enseigner).

L'ensemble des tâches proposées sur une notion à enseigner (scénario) est mis en relation avec la conceptualisation visée, que nous traduisons en termes de tâches et de connaissances, outils et objets, caractérisées par un certain niveau de formalisation, de rigueur et des modes raisonnement attendus et à organiser dans un ensemble de connaissances « déjà-là » (cf. niveau de conceptualisation, dans le prolongement des champs conceptuels de Vergnaud). On distingue les connaissances qui devront être disponibles, utilisées sans indication, des autres, juste mobilisables.

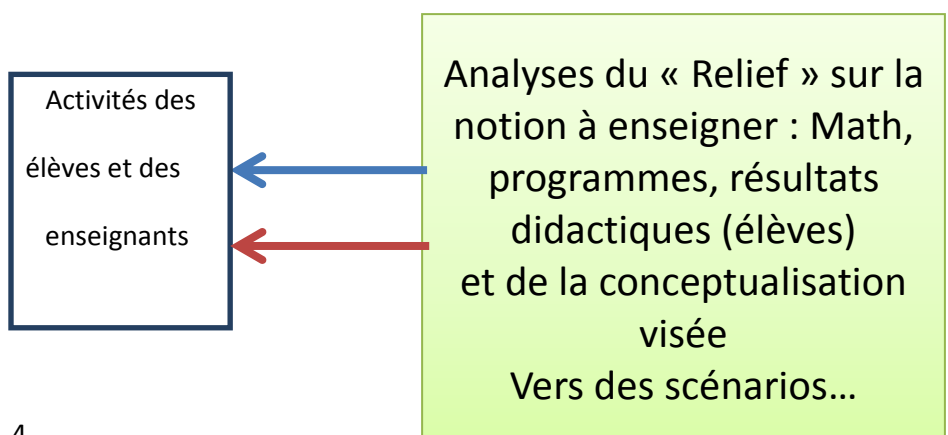

Les activités de l'enseignant incluent la préparation avant la classe (et l'anticipation associée) et l'improvisation durant la classe (par rapport au projet prévu). Mais ces descriptions ne suffisent pas à comprendre les choix réels des enseignants, qui mettent en jeu à la fois les conceptions personnelles (représentations du métier, expériences, connaissances), les contraintes institutionnelles (programmes, horaires notamment, mais aussi inspection), et sociales (côté élèves et côté collectif des collègues et de l'établissement) (cf. double approche didactique et ergonomique, inscrite dans la TA, avec les descriptions en termes de composantes spécifiques à imbriquer *in fine*).

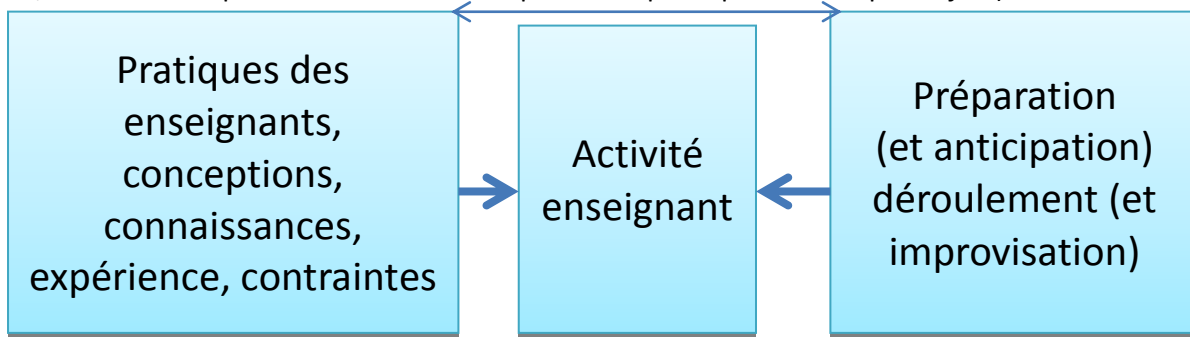

Pour analyser les pratiques des enseignants de manière plus générale, nous ajoutons à ces descriptions trois niveaux d'organisation imbriqués (global – celui des projets, local – celui de la classe, micro – celui des automatismes et routines) qui permettent aussi de les approcher – par exemple chez les enseignants débutants, les niveaux global et micro ne sont pas encore développés.

D'autres facteurs, notamment des aspects affectifs, ou ceux liés à la confiance en soi pour les élèves par exemple, sont peu étudiés par tous les chercheurs concernés. Cependant les origines sociales (voire culturelles) sont prises en compte explicitement dans un certain nombre de recherches, notamment dans le primaire, donnant lieu à des travaux spécifiques sur les activités des élèves et des enseignants dans les classes ou les écoles « défavorisées ».  On peut se demander si la prise en compte de tels facteurs n'amènerait pas à introduire pour les activités des élèves également des niveaux d'organisation, global (posture vis-à-vis de l'école par exemple, projet, rapport au savoir), local (participation en classe), micro (automatismes en termes d'écoute par exemple).

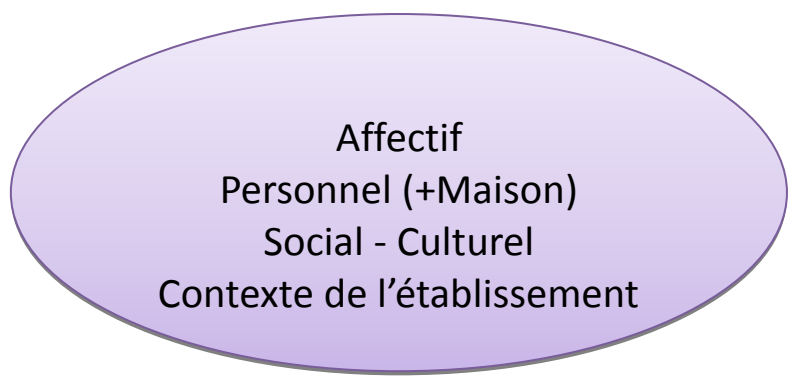

**Méthodologie et emprunts théoriques (cf deuxième schéma à la fin)**

Nous avons ainsi mis au point un certain nombre d'analyses : des tâches (avec différentes adaptations des connaissances par exemple, induisant différentes activités attendues, prolongeant là encore certains travaux de Vergnaud) et des déroulements (avec le pointage de différentes formes de travail, de commentaires des enseignants, d'aides, de rapprochements avec les acquis ou les questions des élèves). Il s'agit de nous permettre d'étudier, en mettant en regard ce qui est attendu et ce qui se passe en classe, les activités possibles[1] des élèves, éventuellement différentes selon les élèves. Au moment d'exposition des connaissances, il s'agit de nous donner des moyens de repérer comment l'enseignant organise en classe les dialectiques du général au contextualisé constitutives des apprentissages visés.

Cela met en jeu des outils spécifiques, adaptés aux mathématiques et au contexte scolaire, dont certains sont déjà utilisés largement en didactique des mathématiques, quel que soit le cadre de référence. D'autres sont encore en développement, notamment du côté du langage, ou de l'intégration des logiciels par exemple. Ces outils sont en partie inspirés[2] par des théories générales d'apprentissage ou de développement que nous avons en quelque sorte opérationnalisées, compte tenu et de nos besoins et de notre domaine particulier.

Citons par exemple les théories de Piaget, en ce qui concerne l'importance des processus de déséquilibre-rééquilibration : ils sont traduits notamment en termes de changements de cadres ou registres (à traquer ou à provoquer). En ce qui concerne l'importance accordée aux phases d'assimilation, d'accommodation et d'abstraction, cela se traduit en particulier par la distinction entre les tâches simples (immédiates), ou plus complexes, comportant des adaptations attendues des connaissances des élèves, à leur charge. Cela amène aussi à repérer les généralisations à la charge de l'enseignant. Ces tâches complexes induisent des activités (supposées) riches, variées, contribuant aux acquisitions visées, qui peuvent porter sur la reconnaissance des connaissances en jeu, et/ou sur l'organisation des résolutions, et/ou sur les traitements à effectuer.
La recherche de tâches où les élèves peuvent travailler seuls, avec une possibilité d'autocontrôle, avant l'introduction formelle et générale d'une définition ou d'une propriété qui les décontextualisent, est aussi issue d'une transposition des théories piagétiennes (particulièrement développée en TSD).

L'analyse des mots utilisés, des commentaires ajoutés (le « méta ») et plus généralement des interactions pendant les déroulements met en jeu une inspiration vygotskienne, centrée sur la qualité des médiations construites dans la classe.
Un élément clef en est le repérage, qui peut être différentiel, des connaissances des élèves déjà-là (ou presque) comme points d'appui pour l'enseignant (cf. opérationnalisation de la notion de ZPD) – ce qui nous inscrit à la suite de Bruner.
En particulier les activités a maxima qualifient des activités que seuls certains élèves font d'emblée. Quand elles ne sont pas reprises intégralement par tous les élèves, ce qui est fréquent, cela en laisse certains hors du champ travaillé par les autres (au moins en partie). Les activités a minima, en revanche, faites par tous, ne sont souvent pas les mêmes que les activités initialement proposées, elles ont souvent été « réduites » en termes de connaissances à mettre en fonctionnement par des interventions de l'enseignant. Leur détection contribue à notre compréhension des activités possibles des élèves.

---
[1] Si ce n'est effectives, hors de portée.
[2] Convertis à partir de , transposés, traduits…

Les proximités en acte caractérisent en particulier les tentatives ou les occasions manquées de rapprochement, repérées par les chercheurs, que l'enseignant opère entre ce qui se fait en classe et ce qu'il veut introduire. Cela participe à l'opérationnalisation pour la classe de mathématiques la notion de ZPD. En ce qui concerne plus précisément les dynamiques entre le général et le contextualisé dans la classe, nous distinguons dans le discours de l'enseignant les proximités discursives ascendantes, qui s'appuient sur les exercices traités par les élèves pour commenter les généralisations visées, les proximités descendantes qui au contraire explicitent la manière d'utiliser le général dans des cas particuliers et les proximités horizontales qui correspondent à des répétitions ou des illustrations sans changement de niveau de généralité du discours en jeu.

**Du local au global**

On peut noter que le facteur « temps » semble absent de nos schémas. De fait les analyses locales, à l'échelle de la séance, ou de quelques séances, et de quelques enseignants, sont plus nombreuses dans nos recherches que les analyses globales.

Pour les enseignants l'hypothèse forte de stabilité[3] des pratiques des enseignants expérimentés, que nous avons développée (et un peu démontrée, cf. double approche) permet une extension des résultats locaux à la détermination de logiques d'action par exemple, pour des enseignants donnés, combinant les 5 composantes rendant compte de la complexité des pratiques. Rappelons qu'il s'agit de spécifier, pour mieux les réimbriquer ensuite, les choix de contenus et de déroulements observables (composantes cognitive et médiative), en tenant compte des contraintes institutionnelles et sociales (composantes du même nom), tout en faisant une place au personnel (composante personnelle). La mise en regard de pratiques de plusieurs enseignants amène à dépasser ces descriptions en établissant des « palettes de possibles », permettant de proposer une certaine variabilité des pratiques, à l'intersection des pratiques observées et des qualités didactiques supposées de scénarios supposés « robustes ».

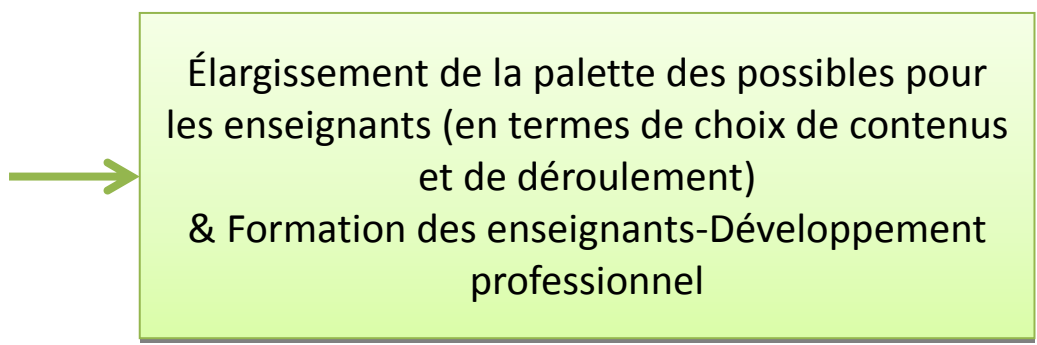

Élargissement de la palette des possibles pour les enseignants (en termes de choix de contenus et de déroulement)
& Formation des enseignants-Développement professionnel

---

[3] Tout particulièrement en ce qui concerne la composante médiative pour des enseignants du secondaire hors évaluation.

Nous avons même prolongé notre réflexion aux formations, en élargissant la notion de ZPD au développement des pratiques (ZPDP, P comme pratiques ou professionnel) et en élaborant des scénarios de formation mettant fortement en jeu un travail collectif sur des pratiques effectives, permettant l'installation de questionnements, de prises de conscience, et de discussions sur les palettes de possibles en termes de choix de contenus et de déroulements en classe. Cela a été l'occasion de travailler aussi sur la transposition de nos outils de recherche à destination des formateurs et des enseignants.

S'agissant des élèves la remontée à du global est plus problématique. Nous ne disposons que d'hypothèses sur la qualité supposée des scénarios pour les apprentissages, et sur l'importance de la récurrence des modes d'enseignement sur les apprentissages provoqués (cf. sociologie).

**Des résultats parmi d'autres**

En termes de résultats, on peut citer rapidement l'obtention d'un certain nombre de palettes de pratiques, analysées, sur différents domaines mathématiques (en primaire sur le numérique ou le géométrique par exemple, dans le secondaire sur l'algébrique ou les fonctions ou les limites) et dans des environnements technologiques.

Un certain nombre d'activités d'introduction, voire de scénarios robustes (notamment sur la symétrie orthogonale) ont été proposés.

Les analyses fines de déroulement, en relation avec celles des activités attendues, ont mis à jour des difficultés d'élèves, quelquefois inattendues, souvent récurrentes. Certaines sont liées en particulier à des implicites que le professeur ne repère pas toujours dans les activités attendues – ce sont même souvent des élèves qui l'amènent à expliciter, notamment si la classe est suffisamment hétérogène. On a ainsi montré, par exemple, que des changements de points de vue, très fréquents en géométrie, ne sont pas toujours intelligibles d'emblée par certains élèves, qui restent plus facilement dans leur seul point de vue initial. Or l'explicitation de leur nécessité ou de leur signification n'en est pas toujours faite, d'autant que certains élèves n'y voient pas de difficultés. Par exemple le passage, dans un exercice, de la présence d'angles droits, à leur traduction nécessaire en droites perpendiculaires peut être passé sous silence, rendant mystérieuse la résolution. D'autres implicites, ne donnant pas lieu à proximités, ont été repérés, par exemple, sur le lien courbe/fonction et le transfert, attendu mais difficilement formalisable, entre les propriétés visuelles globales des courbes et celles algébriques des fonctions représentées.

Nos recherches peuvent aussi amener à des appréciations critiques des programmes ou des instructions : ainsi peut-on analyser qu'il peut exister des tensions entre des exigences de rigueur hétérogènes à différents moments d'un même programme, en particulier liées au travail sur logiciel *vs* papier-crayon ; ou encore que le fait de faire travailler les élèves sur des situations ouvertes peut rendre extrêmement difficile la synthèse qu'on attend et la mise en évidence des connaissances générales à retenir, vu la dispersion des procédures mises en œuvre (dont certaines peuvent même être à la fois efficaces et très « pauvres » mathématiquement, notamment par utilisation d'un bon logiciel).

Un ensemble de livres et d'articles publiés (cf. bibliographie) présentent ces résultats.

**Difficultés, questionnements**

Les difficultés de nos recherches, outre le passage au global, tiennent en particulier au temps qu'elles prennent, à la masse de données à recueillir qu'elles imposent, et à tous les paramètres variables en jeu. D'une année à l'autre les classes changent, les programmes souvent aussi, et les comparaisons sont délicates.

De plus l'emprunt de la notion de ZPD doit être précisé, dans la mesure où la notion est individuelle alors que nous l'utilisons dans le contexte d'une classe.

De façon générale, les validations de ce type de recherches ne sont pas simples, quels que soient les cadres utilisés, d'ailleurs. Les rapports chercheurs – enseignants, indispensables notamment pendant les phases expérimentales, ne sont pas toujours transparents non plus et nécessitent aussi d'être éclaircis, y compris pour la validation.

**Développements récents et perspectives**

Les nouveautés concernent la mise au point d'outils permettant de davantage cibler la distance entre ce que font et/ou savent les élèves et les interventions des enseignants (proximités-en-acte). Mais aussi les études spécifiques des moments d'exposition des connaissances ont progressé, à travers la mise au point d'analyses en termes de proximités discursives, pour apprécier notamment des occasions de proximités possibles, voire manquées, entre ce qui est général et énoncé par l'enseignant et ce que les élèves ont fait sur des cas particuliers ou savent déjà. Les analyses spécifiques des activités avec des outils technologiques permettent d'accéder à ce qui est nouveau en termes de travail sur ces instruments, aussi bien pour l'élève que pour l'enseignant, et de fournir les moyens pour davantage en tenir compte. Viennent se rajouter aussi les pratiques d'évaluations formatives et sommatives, les recherches collaboratives et l'éclaircissement des rôles respectifs dans la collaboration, la formation et l'accompagnement des professeurs d'école en classes très défavorisées.

Reste cependant à expliquer ce qu'apporte ce cadrage par rapport aux autres théories fréquemment utilisées en didactique des mathématiques.

Donnant *une place aux sujets dans leur singularité*, ce cadrage est spécifiquement adapté aux études de ce qui se passe en classe, que les pratiques soient ordinaires ou non d'ailleurs.

D'autres cadres, complémentaires, comme la TSD, se sont d'abord intéressés particulièrement à la conception de situations potentiellement favorables aux apprentissages, dont l'implémentation doit être expérimentée. Même si aujourd'hui ce cadre est utilisé aussi pour étudier des classes ordinaires, ou mettre au point des ressources, l'étude concerne davantage le potentiel attaché à une situation que son déroulement effectif, alors que l'utilisation de la TA inverse cette pondération. Cela amène à dégager ce qui vient du milieu (présent dans la situation indépendamment de l'enseignant), mais aussi les variables didactiques à la disposition de l'enseignant qui lui permettent de jouer sur les activités possibles des élèves. Le contrat didactique sert à spécifier les attentes, explicites ou non, du maître et des élèves les uns envers les autres et sa mise en évidence peut permettre de comprendre ce qui pourrait fausser ou renforcer les jeux dans lesquels sont engagés les élèves. Cela induit une conception des acteurs comme sujets génériques, assurant une fonction (élève, enseignant) plutôt que comme sujets singuliers, acteurs agissant, comme c'est le cas dans notre théorie.

La TAD quant à elle s'intéresse pour partie à travailler sur des visions globales du système éducatif et de la profession (pour les mathématiques), dégagées à partir des contraintes et des normes en vigueur. Les analyses mathématiques de référence s'appuient sur la mise en évidence des praxéologies en jeu, qui vont des types de tâches et de techniques aux technologies en usage et aux théories dans lesquelles elles s'inscrivent. Par ailleurs les phénomènes repérés s'inscrivent dans divers niveaux de détermination, de la classe à la société. Cela induit une conception des acteurs comme sujets assujettis à une institution donnée, et pas, là non plus, comme acteurs singuliers.

Enfin, il est important d'avoir conscience du fait que les théories, quelles qu'elles soient, ne sont pas des carcans mais des garants : il est important de les utiliser pour garantir une certaine cohérence du découpage de la réalité mis en jeu, ce qui est regardé comme variables ou comme paramètres (l'affectif par exemple), mais il est important aussi de repérer ce qui est inattendu, voire de savoir transformer ce qui apparait d'abord comme un couac, un "bruit", ... en une source productive de nouveau (la notion de FUG a été imaginée, par exemple, parce qu'on n'arrivait pas à trouver de «bonne » situation d'introduction aux concepts en jeu en termes de dialectique outil/objet).

De même, si le recueil de données doit être adapté aux cadres théoriques, cela peut produire quand même peut-être de l'inattendu, à savoir repérer… Cela nécessite d'analyser a priori ce qui va être recueilli et d'envisager la portée et les limites de ce qu'on va obtenir. De manière à ne pas chercher des décimales d'un nombre dont le chiffre des unités n'est pas sûr !

**Et dans d'autres domaines ? Une question ouverte…** Vers une théorie de l'activité en didactique des mathématiques et des sciences (**TADMS) ?**

Les approches présentées ont été développées dans le cadre des mathématiques. La question de leur pertinence pour d'autres domaines d'enseignement/apprentissage est posée, et a même déjà reçu des réponses positives dans certaines recherches, notamment sur les pratiques enseignantes (Kermen, 2016), à condition de prendre en compte les différences ou les spécificités épistémologiques des disciplines où on les étend.

De plus, tout comme la notion de champ conceptuel a été élaborée par Vergnaud pour les mathématiques - de l'école élémentaire - et utilisée largement ensuite dans des recherches sur l'enseignement/apprentissage en sciences, la notion de relief semble pertinente pour d'autres domaines de sciences. Mais il est possible qu'il y ait des éléments que nous n'analysons pas en mathématiques et qui seraient à prendre en compte par exemple en physique (ne serait-ce que la "physique de sens commun", qui ne concerne pas seulement les objets d'enseignement de l'élémentaire). Plus généralement, le rapport avec le réel des expérimentations et des théories en physique ou en chimie (différant d'ailleurs selon les disciplines) n'a pas son pendant dans les mathématiques sur lesquelles nous avons travaillé.

## Un schéma global, sans et avec les emprunts théoriques

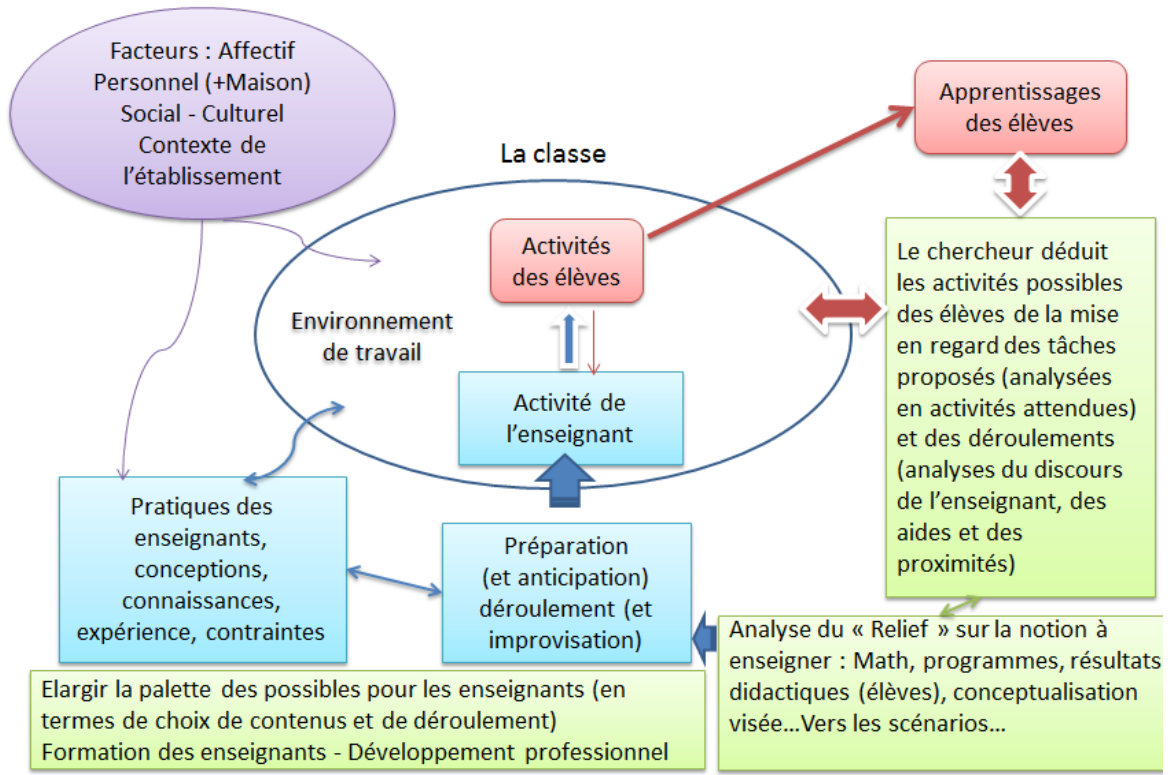

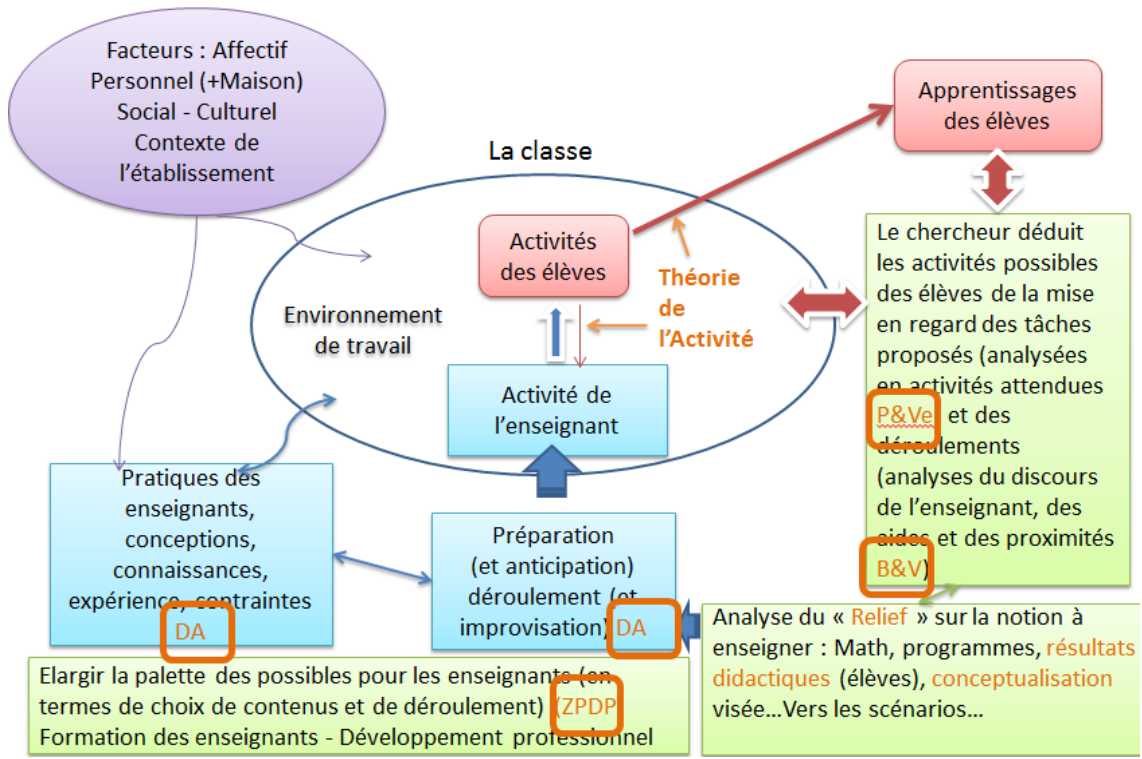